 \documentstyle[prl,aps]{revtex}  %% Journal Style.
\begin{document}
\draft
%\preprint{\today}

%% Include following two lines for Journal Style

 \twocolumn[\hsize\textwidth\columnwidth\hsize  %**Journal
 \csname @twocolumnfalse\endcsname              %**Journal

\title{Superdeformed Band in $^{36}$Ar
       Described by Projected Shell Model}
\author{Gui-Lu Long$^{1,2}$ and Yang Sun$^{1,3,4}$}
\address{
$^1$Department of Physics, Tsinghua University, Beijing 100084, P.R. China\\
$^2$Institute of Theoretical Physics, Chinese Academy of Sciences,
Beijing 100080, P.R. China\\
$^3$Department of Physics and Astronomy, University of Tennessee,
Knoxville, Tennessee 37996\\
$^4$Department of Physics, Xuzhou Normal University,
Xuzhou, Jiangsu 221009, P.R. China} 

%\date{\today}
\maketitle

\begin{abstract}
The projected shell model implements shell model configuration mixing 
in the projected deformed basis. 
Our analysis on the recently observed superdeformed band in $^{36}$Ar suggests 
that the
neutron and proton 2-quasiparticle and the 4-quasiparticle bands 
cross the superdeformed ground band at the same angular momentum.
This constitutes a picture of band disturbance
in which the first and the second band-crossing,
commonly seen at separate rotation frequencies in heavy nuclei,
occur simultaneously.
We also attempt to understand the assumptions 
of two previous theoretical calculations 
which interpreted this band. 
Electromagnetic properties of the band are predicted. 
\end{abstract}


%%  Include the following line for Journal style
 ]  %**Journal

\narrowtext
%\newpage
\bigskip

The topic of superdeformation has been at the forefront
of nuclear structure
physics since the observation of the first superdeformed (SD) band in $^{152}$Dy
\cite{Dy152}. Today, superdeformation at high spin is 
not an isolated phenomenon, but instead is observed 
across the nuclear periodic table \cite{SD-Data},   
and its microscopic foundation has been firmly established. 
However, with the recent observation of the SD band in $^{36}$Ar \cite{Ar36}, 
it is astonishing that the quantum shell effects can stabilize the
system at superdeformation in a nuclear system with such 
few particles (here $N=Z=18$). 

These new data have a large impact on theories, 
as they provide an ideal test case for nuclear structure models. 
The $^{36}$Ar SD data presented in Ref. \cite{Ar36}  
were discussed  
by two theoretical calculations,  
the Cranked Nilsson-Strutinsky (CNS) model \cite{tod} 
and the spherical shell model (SM) \cite{SM}.  
The fact that these models can give a complementary description
for the SD band in $^{36}$Ar indicates 
that they both are appropriate approaches.
Nevertheless, certain assumptions were made in both calculations. 
On the one hand, for a feasible SM calculation, 
the $1d_{5/2}$ orbital had to be excluded from the shell model space.
It is known that 
in the deformed single-particle picture for the present SD  
minimum, the orbital $K={5\over 2}$ of $1d_{5/2}$ lies very
close to the Fermi levels, and it is expected that this orbital has  
strong correlation with other orbitals 
and contributes to the collective motion.  
It is therefore not obvious that 
excluding $1d_{5/2}$ is a proper approximation.
On the other hand, no such exclusion is needed in the CNS calculations. 
However, pairing correlations were completely 
neglected in the CNS 
although there has been no indication that pairing plays a minor role 
in this nucleus. 

The projected shell model (PSM) \cite{review} is a shell model
truncated in the Nilsson single-particle basis, with
pairing correlation incorporated into the basis by a BCS
calculation for the Nilsson states.
More precisely, the truncation is first implemented in the
multi-quasiparticle (qp) basis with respect to the deformed BCS vacuum
$\left|0\right>$ (see Eq.\ (1) below); then the violation of rotational
symmetry is removed by
angular momentum projection \cite{RS80} to
form a shell model basis in the laboratory frame;
finally a shell model Hamiltonian is
diagonalized in this projected space.
Thus, the PSM has the main advantages of 
mean-field theories because it can easily build in the model the most
important nuclear correlations.
It furthermore solves the problem fully quantum mechanically
and provides a good approximation to the exact shell model solution. 
In fact, besides systematic reproductions of energy spectra and electromagnetic 
transitions in normally deformed nuclei \cite{review}, 
it has been shown that 
the SD bands in the $A\sim 190$ \cite{SD190}, 
$A\sim 130$ \cite{SD130} and $A\sim 60$ \cite{SD60} mass regions 
can be successfully described by the PSM. 

It is clear that the PSM lies conceptually
between the two approaches of the CNS and SM in \cite{Ar36}. 
In this paper, we use the PSM 
to analyze the new $^{36}$Ar SD data
and show that it 
gives comparable results to the SM in the spectrum calculation.
The observed band disturbance in
this SD band \cite{Ar36} can be
understood in the PSM framework as simultaneous band-crossings  
among 
the SD ground band (g-band), 2-qp, and 
4-qp bands at the same angular momentum. 
These 2- and 4-qp bands are based on the qausiparticles of the 
$1f_{7/2}$ subshell. 
Quantities such as B(E2), g-factor, and pairing gap are also studied, 
to understand the assumptions
in the CNS and SM calculations mentioned above. 
 
In the present PSM calculation, particles in  
three major shells ($N=1,2,3$) for both neutron and proton are
activated so that the Fermi level lies roughly 
in the middle of the deformed single-particle states
at deformation $\varepsilon_2=0.42$. 
The shell model space includes the 
0-, 2- and 4-qp states:
\begin{equation}
\left|\phi \right>_\kappa  = \left\{\left|0 \right>, \ 
\alpha^\dagger_{n_i} \alpha^\dagger_{n_j} \left|0 \right>,\
\alpha^\dagger_{p_k} \alpha^\dagger_{p_l} \left|0 \right>,\
\alpha^\dagger_{n_i} \alpha^\dagger_{n_j} \alpha^\dagger_{p_k}
\alpha^\dagger_{p_l} \left|0 \right> \right\} ,
\label{baset}
\end{equation}
where $\alpha^\dagger$ is the creation operator for a qp and the
index $n$ ($p$) denotes neutrons (protons).
The projected qp-vacuum $\left|0 \right>$ 
corresponds to the SD g-band,
whereas the projected 2- and 4-qp states to 2- and 4-qp bands, respectively. 
The 2- and 4-qp states are selected so that the low-lying states
for each kind of configuration should be included.  
If all multi-qp states were considered in Eq. (\ref{baset}), 
one would obtain the full shell model space generated by 
particles of the three major shells. 

As in the usual PSM calculations, we employ the Hamiltonian
\cite{review}
\begin{equation}
\hat H = \hat H_0 - {1 \over 2} \chi \sum_\mu \hat Q^\dagger_\mu
\hat Q^{}_\mu - G_M \hat P^\dagger \hat P - G_Q \sum_\mu \hat
P^\dagger_\mu\hat P^{}_\mu,
\label{hamham}
\end{equation}
where $\hat H_0$ is the spherical single-particle Hamiltonian which
contains a proper spin-orbit force, whose strengths (i.e.
the Nilsson parameters $\kappa$ and $\mu$) are taken from Ref.
\cite{tod}.
The second term in the Hamiltonian is the Q-Q interaction and the last
two terms are the monopole and quadrupole pairing interactions,
respectively. 
The interaction strengths are determined as follows: the
Q-Q interaction strength $\chi$ is adjusted by the self-consistent
relation such that the input quadrupole deformation $\varepsilon_2$ and
the one resulting from the HFB procedure coincide with each other
\cite{review}. The monopole pairing strength $G_M$ is taken to be
$G_M=\left[19.6-15.7(N-Z)/A\right]/A$ for neutrons and $G_M=19.6/A$ for
protons. This choice
of $G_M$ seems to be appropriate for the single-particle space employed
in the present calculation in which the major shells $N=1,2,3$ are
included. Finally, the quadrupole pairing
strength $G_Q$ is assumed to be proportional to $G_M$, the
proportionality constant
being fixed to 0.20 in the present work.

The eigenvalue equation of the PSM for a given spin $I$ takes the
form \cite{review}
\begin{equation}
\sum_{\kappa'}\left\{H^I_{\kappa\kappa'}-E^IN^I_{\kappa\kappa'}\right\}
F^I_{\kappa'}=0.
\label{psmeq}
\end{equation}
The expectation value of the Hamiltonian with respect to a ``rotational
band $\kappa$'' $H^I_{\kappa\kappa}/N^I_{\kappa\kappa}$ defines  
a band energy, and when plotted as functions of spin $I$, we call
it a band diagram \cite{review}. 
A band diagram displays bands 
of various configurations before they are mixed by the
diagonalization procedure of Eq. (\ref{psmeq}). 
Irregularity in a spectrum may appear if a
band is crossed by another one(s) at certain spin. 

For the present problem, the eigenvalue equation is solved for different spins
up to $I=16$. This is the highest spin state of 
the SD band if the maximum
spin contributed from the single particles is simply counted \cite{Ar36}.   
In the context of projection, spin distribution in
each basis state of Eq. (\ref{baset}) is given by  
${_\kappa\langle} \phi| \hat P^I_{K_\kappa K_\kappa} |\phi 
\rangle_\kappa$ \cite{bhatt},
where $\hat P^I_{K_\kappa K_\kappa}$ is the projection operator \cite{RS80}. 
We have computed this quantity for each basis state and found that 
they approach zero for spins $I > 16$. In other words,
one cannot find spin larger than 16 in the mean field states
in the present problem.
This is band termination in the language of angular momentum projection. 

Close to the neutron and proton Fermi levels of $^{36}$Ar at deformation
$\varepsilon_2=0.42$, there are four single-particle orbitals: 
$K={5\over 2}$ of $1d_{5/2}$ and $K={1\over 2}$ of $2s_{1/2}$ in
the $N=3$ shell, and 
$K={1\over 2}$ and ${3\over 2}$ of $1f_{7/2}$ in
the $N=4$ shell. 
Thus, bands based on these orbitals are important for determining
the high-spin properties of the low-lying states. 
In Fig. 1, the band diagram is shown. Different
configurations are distinguished by different types of lines, and the
filled circles represent the yrast states obtained after the
configuration mixing. There are about 20 bands in the calculation,
but only representative ones are displayed for discussion.
Note that for the 2-qp bands, one curve represents two bands (a neutron band
and a proton band) because they nearly coincide with each other for the entire
spin region. 

Among the 2-qp bands which start at energies
of 5 -- 6 MeV, one of them (dotted curve) 
consists of two $1f_{7/2}$ quasiparticles 
with $K={1\over 2}$ and $3\over 2$
coupled to total $K=1$. 
It shows a unique behavior as a function of
spin. As spin increases, it goes down first but turns up at spin $I=4$. This
behavior has its origin in the spin alignment of a decoupled band as
intensively discussed in Ref. \cite{review}. Because of this, it can
cross the g-band at about $I=10$.
On the other hand, there is another kind of 2-qp band (long-dashed curve,
based on the coupling of $K={5\over 2}$ of $1d_{5/2}$ 
and $K={1\over 2}$ of $2s_{1/2}$) 
that shows a very different behavior: 
it goes up nearly parallel to the g-band,
and has a very similar form as the g-band. 
This coupled band can never enter into the yrast region, 
thus playing a
negligible role for the yrast band structure.  

We have examined the other multi-qp states consisting of the $1d_{5/2}$
particles, such as the 2-qp state coupling $K={3\over 2}$ and $5\over 2$
to $K=1$.
They lie in an even higher energy region, and have
similar rotational behavior as the g-band. 
As far as the yrast energies are concerned, contributions of the $1d_{5/2}$
orbital to the spectrum calculations can therefore be renormalized. 
Influence of the $1d_{5/2}$
orbital on the absolute values of 
quadrupole moment can also be considered through 
the effective charges.  
This may have clarified the question of 
why the SM reproduced the data remarkably 
well even though it excluded the $1d_{5/2}$
orbital in the calculation \cite{Ar36}. 
 
The two decoupled ($K=1$) 2-qp bands can be combined to a ($K=2$) 4-qp band
which represents simultaneously broken neutron and proton pairs. In Fig.
1, this 4-qp band (solid curve) 
also exhibits a decoupling behavior,
and therefore, the 4-qp band can dive into the yrast region as well.  
It is interesting to see that bands from the three different configurations
(0-, 2-, and 4-qp) cross at the same place near spin $I=10$. 
This is in contrast to the common band-crossing picture leading to
back-bendings in moment of inertia \cite{RS80}. In the usual picture,  
one distinguishes two kinds of 
band-crossings: the first crossing between the g-band and the
2-qp bands, and the second crossing
between the 2-qp and the 4-qp bands. 
They cause the first and the second back-bending in moment of inertia, 
typically seen in the rare
earth nuclei at spin $I \sim 12$ and $\sim 24$, respectively 
\cite{review}.  
The band-crossing picture in $N=Z$ 
nuclei in which a 4-qp band crosses directly with the g-band 
was suggested earlier by Sheikh {\it et al.} \cite{Shei90} 
and further elaborated in Ref. \cite{Frau99}.

Thus, we can interpret the band disturbance in $^{36}$Ar as a
consequence of the simultaneous breaking of the $1f_{7/2}$ neutron and
proton pairs.
After the band-crossing, the main component
of the yrast band is from the 4-qp band.
We observe that all the (0-, 2-, and 4-qp) bands
shown in Fig.\ 1 behave similarly at higher
spins: above spin $I=10$,
all bands displayed are approximately
parallel, indicating that they rotate with the same
frequency.

In Fig. 2a, the PSM energy levels are compared with data, and with 
the SM calculations \cite{Ar36} in the $E(I)-E(I-2)$ plot. 
We observe that the PSM can reasonably 
reproduce the data and the results are comparable with those of the SM.
Following the SD band, one sees that the discontinuity 
around spin $I=10$, 
which corresponds to the band-crossing discussed earlier, 
has been reproduced. 
Nevertheless, in contrast to near-perfect
agreement at the low spins, 
the PSM calculation has small deviations from the data at the 
band-crossing region, and for the higher spin states. 
For the $N\sim Z$ nuclei, 
there has been an open question of whether the proton-neutron
pair correlation plays a role in the structure discussions.
It has been shown that with the renormalized pairing interactions of  
the like-nucleons in an effective
Hamiltoinan, one can account for the $T=1$ part of the proton-neutron pairing 
\cite{Frau99}.
However, whether the renormalization is sufficient for the complex region
that exhibits the phenomenon of band-crossings,
in particular when both neutron and proton pair alignments
occur at the same time, is an interesting question to be investigated.
We note also that the amount of angular momentum gained by the alignment 
is below what one expects from decoupled $f_{7/2}$ pairs. 
Experiment on the neighboring odd-mass nuclei may help us to understand
this issue. 

Fig. 2b and 2c present the calculated B(E2) and g-factor values for the 
$^{36}$Ar SD band. We found that  
the band-crossing does not cause sudden changes around the
crossing spin in either quantity. 
In the B(E2) calculations, 
the effective charges are 0.5e for neutrons and 1.5e for protons, 
which are the same as those used in previous work and in other
shell models \cite{Cr48}. 
We emphasize that employment of different effective charges 
can modify the absolute B(E2) values, but the essential spin dependence 
is determined by the wave functions. 
In Fig. 2b, 
the B(E2) values begin to decrease after spin $I=8$, and a smooth decrease 
is seen for higher spin states. 
At the band termination spin $I=16$, an approximate $40\%$ drop in B(E2) 
(compared to the maximum value at $I=8$) is predicted. 
Our results thus suggest that a considerable collectivity remains even at
the band termination.
In the g-factor calculations, we use for $g_l$ the free values
and for $g_s$ the free values damped by the usual 0.75 factor. 
The results are presented in Fig. 2c.
We observe a smooth increase in the g-factor
from 0.4 at the bandhead to $Z/A$ = 0.5 at $I=8$, and this rotor value
remains thereafter.
The nearly constant g-factor at higher spins indicates a 
cancellation between the proton and the neutron contribution.
To see this clearly, 
we plot two additional curves in Fig. 2c
where the neutron and proton contributions 
are separated. This is done by eliminating the proton (neutron) 
qp states in Eq. (\ref{baset}) 
in the calculation for neutron (proton)
contribution. 
It is now seen that
the proton alignment increases the g-factor, leading a peak at $I=8$, 
whereas the neutron alignment decreases it, causing a valley at the same spin.
The average of the two curves gives the total g-factor that shows a
flat behavior as a function of spin.
This reinforces our previous conclusion about the simultaneous breaking 
of the $1f_{7/2}$ neutron and proton pairs and their combined alignment. 
To test these predictions, 
lifetime measurements for the states in the $^{36}$Ar SD band are required, and 
we hope that recently developed techniques \cite{g-factor} 
can permit the g-factor measurement. 

We finally show the calculated pairing gaps in Fig. 2d, 
in which expectation values of the pair operator are calculated by 
using the PSM wave functions. 
It is found that for this lightest SD nucleus, 
both neutron and proton pairing gaps
are larger than 1 MeV at $I=0$, 
which is a non-negligible value that is of comparable 
size to pairing 
gaps in a heavy, deformed system. 
However, the pairing gaps fall 
quickly as the nucleus rotates. After $I=8$, the falling
continues, and saturates eventually at 
0.3 -- 0.4 MeV.  
This suggests that in order to describe the low-spin spectrum 
properly, pairing and its dynamic evolution are important. 
For the
high spin states, the remaining weak pairing correlation may play a role
in sustaining collectivity. 

In summary, the new experimental data of the SD band in $^{36}$Ar, 
the lightest SD nucleus reported so far, has been described by the PSM. 
The calculated energy levels agreed well with the data, 
as well as with the SM results. 
We may thus conclude 
that the PSM is an efficient shell model truncation scheme 
for the well-deformed light nuclei also, in which the quadrupole collectivity
and pairing correlations dominate the properties. 
Similar conclusions have been drawn in the study of $^{48}$Cr \cite{Cr48}. 
It has been found that in the present case, the 0-, 2-, and 4-qp bands cross 
each other at about spin $I=10$. 
Therefore, the 2-qp configurations do not have a chance 
to play a major role in the structure of the SD yrast band because 
immediately after the band-crossing, 
the 4-qp band dominates the band structure.   
Analysis of the rotational behavior for various bands in the band diagram 
and calculation of the pairing gaps could help us 
understand the assumptions
in the CNS and the SM calculations that were previously
used to interpret the data. 
Electromagnetic properties in this SD band have been studied
with predictions made for the B(E2) and g-factor values. 

Communication with Dr. C. Baktash is acknowledged. 
The authors 
sincerely thank Dr. D.J. Hartley for careful reading the manuscript. 
Y.S. thanks the Department of Physics of Tsinghua University 
for warm hospitality, and for support from
its senior visiting scholar program. 
This work is supported also by the Major State Basic Research Developed Program
Grant No. G2000077400, the China National Natural Science Foundation Grant
No. 19775026, the Fok
Ying Tung Education Foundation, and the
Excellent Young University Teachers' Fund
of Education Ministry of China.

\baselineskip = 14pt
\bibliographystyle{unsrt}

\begin{thebibliography} {99}

\bibitem{Dy152} P.J. Twin {\it et al.}, Phys. Rev. Lett. {\bf 57}, 811 (1986).
\bibitem{SD-Data} X.-L. Han and C.-L. Wu, At. Data Nucl. Data Tables 
 {\bf 73}, 43 (1999); B. Singh, {\it et al.},
 Nucl. Data Sheets, {\bf 78}, 1 (1996).
\bibitem{Ar36} C.E. Svensson {\it et al.}, Phys. Rev. Lett. 
 {\bf 85}, 2693 (2000).
\bibitem{tod} T. Bengtsson and I. Ragnarsson, Nucl. Phys. {\bf A436}, 14
 (1985).
\bibitem{SM} E. Caurier, A.P. Zuker, A. Poves and G. Mart\'\i nez-Pinedo, 
 Phys. Rev. {\bf C50}, 225 (1994).
\bibitem{review} K. Hara and Y. Sun, Int. J. Mod. Phys. {\bf E4}, 637
 (1995).
\bibitem{RS80}
 P. Ring and P. Schuck, {\em The Nuclear Many Body Problem}
 (Springer-Verlag, New York, 1980).
\bibitem{SD190} Y. Sun, J.-y. Zhang and M. Guidry,
 Phys. Rev. Lett. {\bf 78}, 2321 (1997). 
\bibitem{SD130} Y. Sun and M. Guidry, Phys. Rev. {\bf C52}, R2844 (1995).
\bibitem{SD60} Y. Sun, J.-y. Zhang, M. Guidry and C.-L. Wu,
 Phys. Rev. Lett. {\bf 83}, 686 (1999). 
\bibitem{bhatt} K.H. Bhatt, S. Kahane and S. Raman, 
 Phys. Rev. {\bf C61}, 034317 (2000).
\bibitem{Shei90} J.A. Sheikh, N. Rowley, M.A. Nagarajan and H.G. Price,
 Phys. Rev. Lett. {\bf 64}, 376 (1990).
\bibitem{Frau99} S. Frauendorf and J.A. Sheikh, 
 Nucl. Phys. {\bf A645}, 509 (1999).  
\bibitem{Cr48} K. Hara, Y. Sun and T. Mizusaki, 
 Phys. Rev. Lett. {\bf 83}, 1922 (1999).
\bibitem{g-factor} R. Ernst {\it et al.}, Phys. Rev. Lett. {\bf 84}, 
 416 (2000). 

\end{thebibliography}

\begin{figure}
\caption{
Band diagram (bands before configuration mixing) and the yrast band
(the lowest band after configuration mixing, denoted by dots) 
for the superdeformed nucleus $^{36}$Ar.
Only the
important lowest-lying
bands in each configuration are shown. 
}
\label{figure.1}
\end{figure}

\begin{figure}
\caption{a) Transition energies $E(I)-E(I-2)$ along 
the superdeformed yrast band in $^{36}$Ar 
(The experimental data and the SM results are taken from Ref. 
\protect\cite{Ar36}); 
b) calculated B(E2) values; 
c) calculated g-factors;
and d) calculated pairing gaps. 
}
\label{figure.2}
\end{figure}

\end{document}